# Hybrid light-emitting polymer/SiN$_x$ platform for photonic integration

Zeming Fan[1], Zeru Wu[1], Yujie Chen[1,*], Zengkai Shao[1], Yanfeng Zhang[1], Zhiren Qiu[1,*], and Siyuan Yu[1,2]

[1]*State Key Laboratory of Optoelectronic Materials and Technologies, School of Physics, School of Electronics and Information Technology, Sun Yat-sen University, Guangzhou510275, China*
[2]*Photonics Group, Merchant Venturers School of Engineering, University of Bristol, Bristol BS8 1UB, UK*
[*]*Corresponding Emails: chenyj69@mail.sysu.edu.cn (Y. Chen) and stsqzr@mail.sysu.edu.cn (Z. Qiu)*

**Abstract:** Organic semiconductors have potentials for a broad range of applications; however, it is difficult to be integrated with traditional inorganic material to meet the need of further application. Based on low-temperature silicon nitride (SiN$_x$) deposition technique, here we demonstrate a hybrid structure fabricated by directly depositing high-quality SiN$_x$ on organic polymer film Poly[2-(2',5'-bis(2"-ethylhexyloxy)- phenyl) -1,4-phenylene vinylene] (BEHP-PPV). Stacked BEHP-PPV/SiN$_x$ hybrid structures with different periods are obtained and their optical properties are systematically characterized. Moreover, a group of BEHP/PPV embedded SiN$_x$ micro-disk is fabricated and amplification of spontaneous emission (ASE) is observed under optical pumping, further confirming that the gain properties of BEHP/PPV are well preserved. Our technique offers a platform to fabricate organic/inorganic hybrid optical devices compatible with integrated components.

## 1. Introduction

Recent research and progress of organic semiconductor has sprouted interest in the fields of organic LED and LET [1-3], organic lasers [4-8], organic sensors [9], and organic telecommunication devices [10,11]. Organic semiconductors have many novel optoelectronic properties, such as ultra-high photoluminescence efficiency, high gain, and tunability, among their broad fluorescence spectrum, making them suitable as laser materials [12]. Especially, conjugated polymers, one kind of organic semiconductors, own desired features of simple fabrication and lost-cost, while it can be adapted to fabricate devices easily using solution process techniques such as spin-coating and ink-jet printing [13]. The advantageous optical gain properties and fabrication flexibility of conjugated polymer make it appealing to be combined with passive materials to fabricate active components compatible with photonic integrated circuits (PICs) [14], which is an essential issue for building fully on-chip integrated systems. Over the past decades, on-chip organic/inorganic hybrid devices have been demonstrated for different applications, for instance, hybrid organic/inorganic lasers and silicon-polymer modulators [15-17].

Considering material properties and fabrication technique, silicon nitride (SiN$_x$) is a promising material as inorganic component for hybrid devices. SiN$_x$ has been widely used in the complementary metal-oxide-semiconductor (CMOS) industry as electrical and chemical isolation [18]. Furthermore, SiN$_x$ is transparent from visible to near-infrared spectral region, making it a high performance solution for PICs [19-21]. Commonly, SiN$_x$ can be deposited with conventional techniques including low-pressure chemical vapor deposition (LPCVD) and plasma-enhanced chemical vapor deposition (PECVD) [22], but high deposition temperature may conflict with many applications related to materials that are not able to survive in high temperatures of hundreds degrees Celsius [18,23]. Recently, inductively coupled plasma chemical vapor deposition (ICP-CVD) technique has been developed to grow SiN$_x$ film under ultra-low temperature [24-26]. In our previous work, we have reported that a high quality SiN$_x$ film with thickness up to 2 μm can be achieved in a single growth process under ultra-low temperature circumstance [26], which shows great potential for active passive hybrid photonic integration [27, 28].

In this work, we fabricate a hybrid structure consisting of stacked SiN$_x$ layer and BEHP-PPV spin-coating film using ultra-low temperature SiN$_x$ deposition technique in ICP-CVD. Here, we have succeeded to deposit high quality SiN$_x$ with thickness of up to 380 nm directly on spin-coating BEHP-PPV film while optical properties of the hybrid structure have been measured. The results confirm that the activity of BEHP-PPV is well preserved after the whole fabrication process. Moreover, we fabricate multi-layer micro-disks with different radius containing a 367-nm-thick SiN$_x$ layer on BEHP-PPV film on the basis of the hybrid structure, in which amplified spontaneous emission (ASE) is observed during the measurement of the multi-layer micro-disk. Our work provides a new strategy for constructing hybrid light-emitting polymer/SiN$_x$ platform which could be applied in integrated photonics.

## 2. Fabrication of multi-layer polymer/SiN$_x$ structure

Conjugated polymer powder BEHP-PPV is purchased from Sigma-Aldrich and dissolved in toluene in 10 mg/mL proportion. The solution is stirred using a vertex mixer until the BEHP-PPV powder is totally dissolved in toluene. Then it is spin-coated to obtain a film with thickness of about 50 nm on a cleaned and pretreated quartz substrate. After that, the spin-coating film is dried under nitrogen atmosphere at 90 °C for 180 min. A thin SiN$_x$ layer with thickness of around 55 nm is deposited on the spin-coating polymer film by means of inductively coupled plasma chemical vapor deposition (ICP-CVD, Oxford Instrument Plasmalab System 100 ICP180) under 75 °C circumstance and a single period of PPV/SiN$_x$ structure (SPS) is obtained. Repeating the spin-coating and SiN$_x$ deposition process, we also fabricate double period PPV/SiN$_x$ structure (DPS) and triple period PPV/SiN$_x$ structure (TPS). The photograph of cross section of TPS under scanning electron microscope (SEM) [Fig. 1(b)] reveals a smooth interface between BEHP-PPV spin-coating film (green layer) and SiN$_x$ layers (blue layer). The photograph of the SPS, DPS and TPS without any crack exposed in UV light around 380 nm is shown in Fig. 1(c).

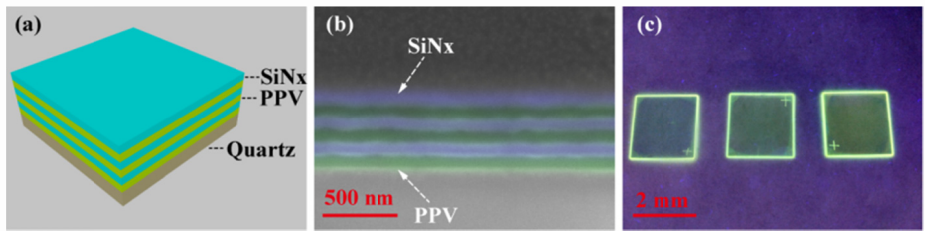

Fig. 1. (a) Schematic illustration of the multi-period structure on quartz substrate (brown layer) composed of spin-coating polymer film (green layer) and SiN$_x$ layer (blue layer). (b) SEM image of the cross section of DPS. (c) Photograph of SPS, DPS, TPS (left to right) under UV light illumination.

The absorption and fluorescence spectra of SPS, DPS, TPS and spin-coating BEHP-PPV film have been measured using an EMCCD (Andor Technology Newton DU907P) attached to a Shamrock SR500i spectrometer. Two obvious peaks around 489 nm and 521 nm can be identified in the fluorescence spectrum of the neat spin-coating BEHP-PPV film [Fig. 2(a)]. As Fig. 2(b) shows, the fluorescence spectrum of DPS keeps consistent with the neat spin-coating BEHP-PPV film. For further characterization, samples are excited by a tunable laser source (Vibrant 355 II, OPOTEK Inc.) which operates at a wavelength of 420 nm with a repetition rate of 10 Hz and a pulse duration of 5 ns. The emission is collected by a fiber coupled multichannel spectrometer platform (Avantes model AVS-DESKTOP-USB2) with 0.34 nm resolution. An obvious vibrational peak around 516 nm can be observed in emission spectrum of DPS [Fig. 2. (b)], which indicates a threshold behavior because peak intensity and full-width at half-maximum (FWHM) of the emission peak evolve as a function of energy density of the pump pulse laser [Fig. 2. (c)]. The emission peak rises and narrows rapidly when the pump energy density exceeds a threshold around 0.083 mJ/cm$^2$ [Fig. 2. (d)]. Moreover, DPS has a larger saturation gain coefficient [Fig. 2. (e)] and lower threshold [vertical dashed line in Fig. 2.(e)] than both SPS and TPS. Gain coefficient of BEHP-PPV film is measured with variable stripe length (VSL) technique [29]. As shown in Fig. 2(e), DPS has a gain saturation coefficient of up to 32 cm$^{-1}$, which implies that gain characteristics of BEHP-PPV is well preserved during the fabrication process. Compared with SPS, the total thickness of the active polymer film in DPS is larger, leading to stronger absorption and emission. However, a spin-coating film usually has a thicker edge, in other words, the quality of SiN$_x$ layer decreases as the number of structure period increases.

Lifetimes of samples are measured with an imaging spectrograph (C5094, HAMAMATSU) and a streak camera system (High speed streak unit M1952, HAMAMATSU), pumped by a Nd:YAG laser (EKSPLAPL2143 B/SS) and an OPG system (EKSPLA PG401SH/DFG2-10) operating at the wavelength of 420 nm with a repetition rate of 10 Hz and a pulse duration of 21 ps. As Fig. 2(f) shows, the lifetime of all samples stabilize to around 23 ps as energy per pulse of the pump laser increases. Among PPV derivative, lifetime is always positively related with photoluminescence (PL) quantum efficiency [30]. Thus, uniformity of the lifetime among SPS, DPS and TPS indicates BEHP-PPV in the hybrid structure still maintain stable after our process flow.

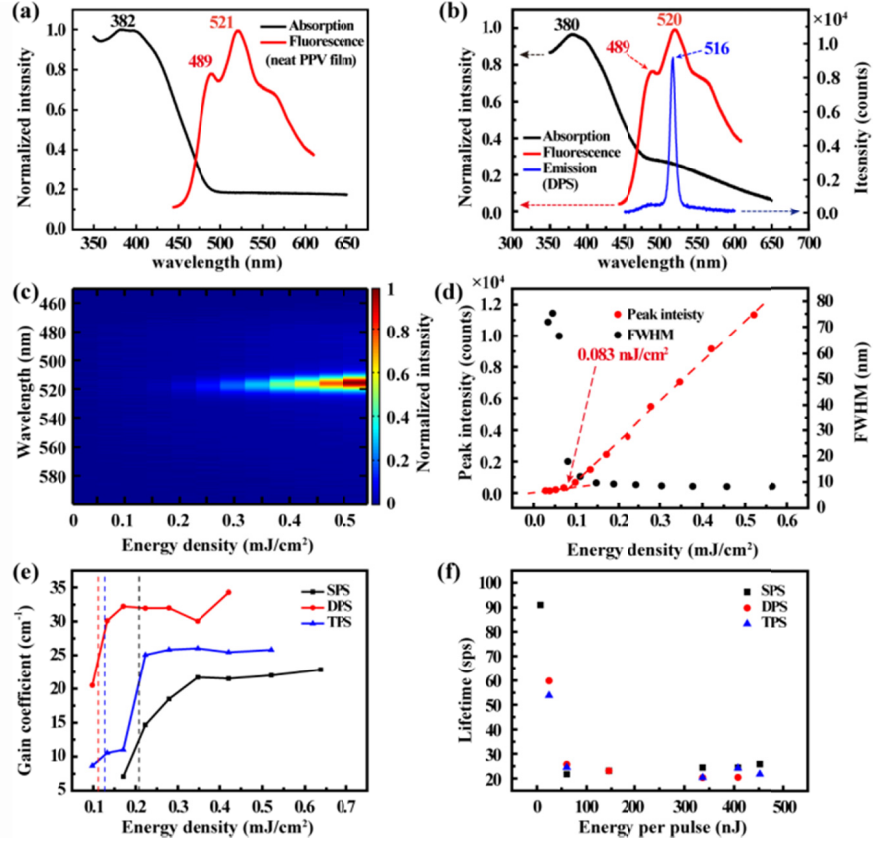

Fig. 2. (a) Fluorescence (red line) and absorption (black line) spectrum of the neat spin-coating BEHP-PPV film on quartz substrate. (b) Fluorescence (red line), absorption (black line), and emission (blue line) spectrum of DPS. (c) Emission spectrum of DPS pumped by a 420-nm, 10-Hz pulse laser with different energy density. (d) Peak intensity and FWHM of emission peak of DPS as a function of the pump energy density. As the pump energy density increases, a rapidly narrowing and rising emission peak around 516 nm can be observed. (e) Gain coefficients of SPS, DPS, and TPS pumped by 10 Hz pulse laser at 420nm with different energy density. The vertical dashed lines indicate the threshold of SPS (black), DPS (red), and TPS (blue), respectively. (f) Lifetimes of SPS, DPS, and TPS pumped by picosecond pulse laser at 420nm with duration of about 21 ps.

## 3. Fabrication of polymer/SiN$_x$ micro-disks

On basis of multi-layer polymer/SiN$_x$ structure, micro-disksare fabricated by opto-lithography and the fabrication process is described as follows. Firstly, a layer of 120 nm-thick BEHP-PPV film is spin-coated on a clean quartz substrate, followed by a depositon of SiN$_x$ layer with thickness of around 370 nm. Then, negative photo-resist AZnLoF2035 (Clariant Corporation) is spin-coated onto the SiN$_x$ layer and exposed with a direct-writing lithography system (Heidelberg InstrumentsuPG501). At last, we fabricate the multi-layer micro-disk with reactive ion etch system (PlamsaPro System 100RIE). Photograph of the multi-layer micro-disk with radius of 76.6 μm under SEM [Fig. 3(b)] shows that the as-fabracated micro-disk has obvious sandwich-like structure with smooth edge and sidewall. The polymer-embedded structure not only provides solid protection for the polymer but also effectively enhances the spatial overlap between gain medium and the resonating modes.

Characterization setup of the multi-layer micro-disk is schematically illustrated in Fig. 3(a). Before the laser pulse focusing onto the backside of the micro-disk, it goes through a cylindrical lens, a Glan-Taylor prism, a polarization beam splitter, a 450-nm-centered band-pass filter and a 50:50 beam splitter. A lens collects the signals emitted from the micro-disk into the fiber-coupled spectrometer while images of the micro-disk are captured by a CMOS camera (Thorlabs, Inc). Here, we fabricate six groups of micro-disks with different radius, 52.3, 76.6, 104.6, 156.9, 209.2, and 261.5 μm. Photograph of the whole sample under UV light illumination [Fig. 4(a)] shows that BEHP-PPV in the sample can still emit strong fluorescence. Fluorescence spectrum of the micro-disk with radius of 261.5 nm under 405 nm light illumination [Fig. 4(b)] keeps consistence with the neat spin-coating BEHP-PPV film [Fig. 2(a)]. Moreover, when pumped by the 420-nm and 10-Hz laser pulses, ASE can be observed in every group of micro-disks. As for the micro-disk with radius of 261.5 μm, there is an emission peak around 516 nm [Fig. 4. (c)]. On the one hand, the emission peak rises rapidly when pump energy density exceeds 6.61 mJ/cm$^2$, indicating a threshold behavior [Fig. 4(d)]. On the other hand, this emission peak blueshifts slightly as the pump density increases, which is related with the self-absorption of BEHP-PPV film. Compared with micro-disk with radius of 52.3 μm, micro-disk with radius 261.5 μm shows more obvious threshold behavior and thus has higher gain.

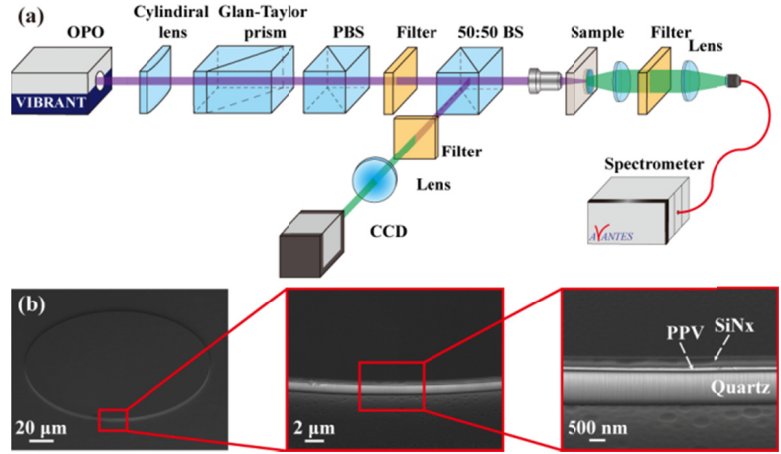

Fig. 3. (a) Schematic picture of the multi-layer micro-disk test setup. Pump pulse laser is focused on the backside of the micro-disk and its emission signal is collected into a fiber-coupled spectrometer. Images of micro-disk are captured by a coaxial camera. (b) SEM photograph of the profile of the multi-layer micro-disk (R=76.6 μm) with smooth sidewall. The thickness of SiN$_x$, PPV and quartz is 367, 123, and 1280 nm, respectively.

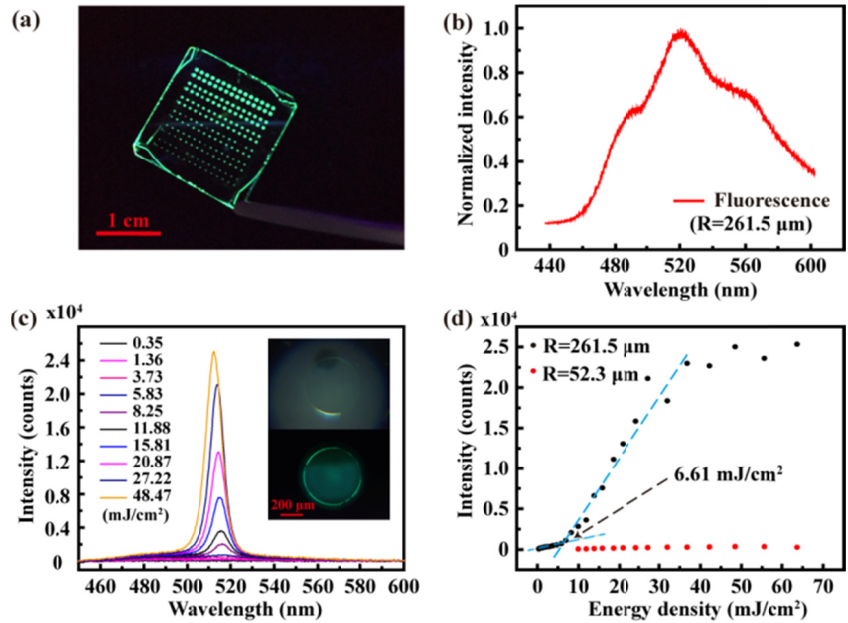

Fig. 4. (a) Photograph of six groups of micro-disks under UV light illumination. (b) Fluorescence spectrum of the micro-disk with radius of 261.5 μm. (c) Emission spectrum of micro-disk with radius of 261.5 μm pumped by a 420-nm and 10-Hz pulse laser with different energy density. The inset picture shows the micro-disk under white light illumination (top) and the pulse laser pump (bottom). (d) Peak intensity of the emission peaks of micro-disks with radius of 261.5 and 52.3 μm, indicating a threshold around 6.61 mJ/cm$^2$ for the 261.5 μm-radius one.

## 4. Conclusion

We have fabricated an organic/inorganic hybrid structure consisting of spin-coating BEHP-PPV film and SiN$_x$ layer. During the whole fabrication processes, activity of BEHP-PPV in the hybrid structure shows no degradation. Typically, saturation gain coefficient of DPS sample can be up to 32 cm$^{-1}$. Then we fabricate six groups of multi-layer hybrid micro-disks with different radius on the basis of such hybrid structure and ASE is observed when the micro-disks are pumped by nanosecond laser pulses. Thus, optical properties of BEHP-PPV still maintain stable in the hybrid structure and the micro-disk, which indicates that the organic/inorganic hybrid structure could provide a new platform for constructing novel hybrid integrated components in many photonic applications.

**Funding.** National Natural Science Foundations of China (51403244, 61323001, 11774437, 11474365); Natural Science Foundation of Guangdong Province (2014A030313104); Science and Technology Program of Guangzhou (201707020017); Fundamental Research Funds for the Central Universities of China (Sun Yat-sen University: 17lgzd06, 16lgjc16).

**References**

1. R. H. Friend, R. W. Gymer, A. B. Holmes, J. H. Burroughes, R. N. Marks, C. Taliani, D. D. C. Bradley, D. A. Dos Santos, J. L. Bredas, M. Logdlund and W. R. Salaneck, "Electroluminescence in conjugated polymers, " Nat. **397**, 121-128 (1999).
2. N. Thejo Kalyani and S. J. Dhoble, "Organic light emitting diodes: Energy saving lighting technology—A review," Renew. Sust. Energ. Rev. **16**, 2696-2723 (2012).
3. R. Capelli, S. Toffanin, G. Generali, H. Usta, A. Facchetti, and M. Muccini, "Organic light-emitting transistors with an efficiency that outperforms the equivalent light-emitting diodes," Nat. Mater. **9**, 496-503 (2010).
4. C. Foucher, B. Guilhabert, A. L. Kanibolotsky, P. J. Skabara, N. Laurand and M. D. Dawson, "RGB and white-emitting organic lasers on flexible glass," Opt. Express **24**, 2273-2280 (2016).
5. Y. Chen, J. Herrnsdorf, B. Guilhabert, A. L. Kanibolotsky, A. R. Mackintosh, Y. Wang, R. A. Pethrick, E. Gu, G. A. Turnbull, and P. J. Skabara, "Laser action in a surface-structured free-standing membrane based on a p-conjugated polymer-composite," Org. Electron. **12**, 62-69 (2011).
6. Y. Wang, X. Yang, H. Li, and C. Sheng, "Bright single-mode random laser from a concentrated solution of π-conjugated polymers," Opt. Lett. **41**, 269-272 (2016).
7. M. Mantoku, M. Ichida, I. Umezu, A. Sugimura, and T. Aoki-Matsumoto, "Lasing in organic mixed-crystal thin films with cavities composed of naturally formed cracks," Opt. Lett. **42**, 1528-1531 (2017).
8. T. Zhai, F. Tong, Y. Wang, X. Wu, S. Li, M. Wang, and X. Zhang, "Polymer lasers assembled by suspending membranes on a distributed feedback grating," Opt. Express **24**, 22028 (2016).
9. D. Tyler McQuade, Anthony E. Pullen and Timothy M. Swager, "Conjugated polymer-based chemical sensors," Chem. Rev. **100**, 2537-5274 (2000).
10. J. Clark, and G. Lanzani, "Organic photonics for communications," Nat. Photon. **4**, 438-446 (2010).
11. Y. Enami, C. T. Derose, D. Mathine, C. Loychik, C. Greenlee, R. A. Norwood, T. D. Kim, J. Luo, Y. Tian, A. K. Y. Jen, and N. Peyghambarian, "Hybrid polymer/sol–gel waveguide modulators with exceptionally large electro–optic coefficients," Nat. Photon. **1**, 180-185 (2007).
12. H. C. Yeh, and M. C. Hsieh, "Tuning the emission color of organic semiconductor films with cholesteric liquid crystals," Opt. Express **25**, 11211-11216 (2017).
13. I. D. W. Samuel and G. A. Turnbull, "Organic Semiconductor Lasers," Chem. Rev. **107**, 1272-1295 (2007).
14. W. Xie, T. Stoferle, G. Raino, T. Aubert, S. Bisschop, Y. Zhu, R. F. Mahrt, P. Geiregat, E. Brainis, Z. Hens, and D. Van Thourhout, "On-chip integrated quantum-dot-silicon-nitride microdisk lasers," Adv. Mater. **29** (2017).
15. Y. Yang, G. A. Turnbull and I. D. W. Samuel, "Hybrid optoelectronics- A polymer laser pumped by a nitride light-emitting diode," Appl. Phys. Lett. **92**, 163306 (2008).
16. T. W. Beahr-Jones and M. J. Hochberg, "Polymer silicon hybrid systems A platform for practical nonlinear optics†," J. Phys. Chem. C **112**, 8085-8090 (2008).
17. M. Hochberg, T. Baehr-Jones, G. Wang, M. Shearn, K. Harvard, J. Luo, B. Chen, Z. Shi, R. Lawson, P. Sullivan, A. K. Jen, L. Dalton, and A. Scherer, "Terahertz all-optical modulation in a silicon-polymer hybrid system," Nat. Mater. **5**, 703-709 (2006).
18. B. De Geyter, K. Komorowska, E. Brainis, P. Emplit, P. Geiregat, A. Hassinen, Z. Hens, and D. Van Thourhout, "From fabrication to mode mapping in silicon nitride microdisks with embedded colloidal quantum dots," Appl. Phys. Lett. **101**, 161101 (2012).
19. S. Romero-Garcia, F. Merget, F. Zhong, H. Finkelstein, and J. Witzens, "Silicon nitride CMOS-compatible platform for integrated photonics applications at visible wavelengths," Opt. Express **21**, 14036 (2013).
20. D. J. Moss, R. Morandotti, A. L. Gaeta, and M. Lipson, "New CMOS-compatible platforms based on silicon nitride and Hydex for nonlinear optics," Nat. Photon. **7**, 597-607 (2013).
21. P. T. Lin, V. Singh, H.-Y. G. Lin, T. Tiwald, L. C. Kimerling, and A. M. Agarwal, "Low-stress silicon nitride platform for mid-infrared broadband and monolithically integrated microphotonics," Adv. Opt. Mater. **1**, 732-739 (2013).
22. J. Yota, J. Hander and A. A. Saleh, "A comparative study on inductively-coupled plasma high-density plasma, plasma-enhenced, and low pressure chemical vapor deposition silicon nitride films," J. Vac. Sci. Technol. **A18**, 372-376 (2000).
23. S. Ueno, Y. Konishi, and K. Azuma, "The structures of highly transparent, water impermeable SiNx films prepared using surface-wave-plasma chemical vapor deposition for organic light-emitting displays," Thin Solid Films **580**, 106-110 (2015).
24. H. Zhou, K. Elgaid, C. Wilkinson, and I. Thayne, "Low-hydrogen-content silicon nitride deposited at room temperature by inductively coupled plasma deposition," Jpn. J. Appl. Phys. **45**, 8838-8892 (2006).

25. D. Dergez, J. Schalko, A. Bittner, and U. Schmid, "Fundamental properties of a-SiNx:H thin films deposited by ICP-PECVD for MEMS applications," Appl. Surf. Sci. **284**, 348-353 (2013).
26. Z. Shao, Y. Chen, H. Chen, Y. Zhang, F. Zhang, J. Jian, Z. Fan, L. Liu, C. Yang, L. Zhou, and S. Yu, "Ultra-low temperature silicon nitride photonic integration platform," Opt. Express **24**, 1865 (2016).
27. Y. Zhu, W. Xie, S. Bisschop, T. Aubert, E. Brainis, P. Geiregat, Z. Hens, and D. Van Thourhout, "On-chip single-mode distributed feedback colloidal quantum dot laser under nanosecond pumping," ACS Photon. **4**, 2446–2452 (2017).
28. P. J. Cegielski, S. Neutzner, C. Porschatis, H. Lerch, J. Bolten, S. Suckow, A. R. S. Kandada, B. Chmielak, A. Petrozza, T. Wahlbrink, and A. L. Giesecke, "Integrated perovskite lasers on a silicon nitride waveguide platform by cost-effective high throughput fabrication," Opt. Express **25**, 13199-13206 (2017).
29. L. D. Negro, P. Bettotti, M. Cazzanelli, D. Pacifici, and L. Pavesi, "Applicability conditions and experimental analysis of the variable stripe length method for gain measurements," Opt. Commun. **229**, 337-348 (2004).
30. C. X. Sheng, M. Tong, S. Singh, and Z. V. Vardeny, "Experimental determination of the charge/neutral branching ratio η in the photoexcitation of π-conjugated polymers by broadband ultrafast spectroscopy," Phy. Rev. B **75**, 085206 (2007).